\documentclass[prl,aps,twocolumn,showpacs,amssymb]{revtex4-1}

\usepackage{dcolumn}
\usepackage{bm}
\usepackage{commath}
\usepackage{graphicx}
\usepackage{setspace}
\usepackage{amsmath}
\usepackage{amssymb}
\usepackage{physics}
\usepackage[normalem]{ulem}
\usepackage{color}

\newcommand{\beginsupplement}{%

        \setcounter{table}{0}
        \renewcommand{\thetable}{S\arabic{table}}%
        \setcounter{figure}{0}
        \renewcommand{\thefigure}{S\arabic{figure}}%
     }

\begin{document}

\preprint{APS/123-QED}

\title{Overdamped Phase Diffusion in hBN Encapsulated Graphene Josephson Junctions}
\author {J. Tang$^{1}$, M.T. Wei$^{2}$, A. Sharma$^{1}$, E. G. Arnault$^{3}$, A. Seredinski$^{4}$, Y. Mehta$^{5}$, K. Watanabe$^{6}$, T. Taniguchi$^{6}$, F. Amet$^{5}$, I. Borzenets$^{1}$$^{*}$}

\affiliation{
$^{1}$Department of Physics, City University of Hong Kong, Kowloon, Hong Kong SAR.
\\$^{2}$Joint Quantum Institute, University of Maryland, MD 20742, USA. 
\\$^{3}$Department of Physics, Duke University, Durham, NC 27708, USA.
\\$^{4}$Department of Sciences, Wentworth Institute of Technology, Boston, MA 02115, USA. 
\\$^{5}$Department of Physics and Astronomy, Appalachian State University, Boone, NC 28607, USA.
\\$^{6}$Advanced Materials Laboratory, National Institute for Materials Science, Tsukuba, 305-0044, Japan.
\\$^{*}$Correspondence should be sent to I.V.B. (email: iborzene@cityu.edu.hk)
 }

\date{\today}

\begin{abstract}
We investigate the zero-bias behavior of Josephson junctions made of encapsulated graphene boron nitride heterostructures in the long ballistic junction regime. For temperatures down to $2.7$K, the junctions appear non-hysteretic with respect to the switching and retrapping currents $I_C$ and $I_R$.  A small non-zero resistance is observed even around zero bias current, and scales with temperature as dictated by the phase diffusion mechanism. By varying the graphene carrier concentration we are able to confirm that the observed phase diffusion mechanism follows the trend for an overdamped Josephson junction. This is in contrast with the majority of graphene-based junctions which are underdamped and shorted by the environment at high frequencies. 
\end{abstract}

\maketitle


Graphene-based superconductor-normal metal-superconductor (SNS) Josephson junctions have been a popular medium of choice for studying the fundamentals\citep{Heersche2007,  Miao1530,  PhysRevB.77.184507,  PhysRevB.79.165436, Phase_Diff, Ke2016,PhysRevLett.107.146605,PhysRevLett.108.097003,Borzenets_Long_Ballistic, BenShalom2016,Larson2020} as well as applications\citep{Borzenets_Splitter, Borzenets_Phonon, Our_Science, Borzenets_T, Lee2017, Park2017} of superconducting devices for more than a decade.  However, the full spectrum and consequences of the interactions between the graphene Josephson junction and the environment have not been fully mapped. For example, the observed critical current $I_C$ of graphene Josephson junctions is consistently suppressed compared to theoretical predictions; leading to postulations that the junctions are severely underdamped\citep{tinkham,  Phase_Diff, Ke2016, Borzenets_Long_Ballistic, BenShalom2016}, despite the relatively low hysteresis between the switching $I_S$ and the retrapping $I_R$ currents. The effect of a junction's environment on its dynamics can be directly investigated by looking at the statistical distribution of the switching current $I_S$ \citep{Borzenets_Long_Ballistic, PhysRevLett.108.097003, PhysRevLett.107.146605, Fulton, Clarke}, or via the measurement of zero-bias resistance arising from the phase diffusion mechanism\citep{tinkham,Phase_Diff,PhysRevB.50.395,PhysRevLett.63.1507,PhysRevB.42.9903,JETP.8.113,JETP.28.1272}. Indeed, often attributed to the large capacitance generated by the bonding pads and leads, previous works have shown that the vast majority of graphene-based JJs are underdamped\citep{Phase_Diff, Ke2016, BenShalom2016, Borzenets_Long_Ballistic}. 

In this work we report on Josephson junctions made from hexagonal boron-nitride (hBN) encapsulated graphene with Molybdenum-Rhenium (MoRe) alloy superconducting contacts \citep{Wang614, Borzenets_T, Our_Science}. These devices are governed by ballistic electron transport and have been found to be in the intermediate to long-junction regime\citep{Borzenets_Long_Ballistic} (see supplementary). Here, the MoRe contacts terminate shortly after the active region and are connected to the bonding pads via thin Gold leads. Moreover, the MoRe-Au interface seems to exhibit a significant contact resistance. The junctions were measured at temperatures between $2.7$ and $7$K, where a clear phase diffusion governed, zero-bias resistance can be observed\citep{tinkham,Phase_Diff,PhysRevB.50.395,PhysRevLett.63.1507,PhysRevB.42.9903,JETP.8.113,JETP.28.1272}.  However, for these devices, when changing the carrier concentration via the back gate, the zero-bias resistance follows the trend expected for overdamped junctions\citep{Phase_Diff, JETP.28.1272,Ambegaokar}. Thus, we conclude that we have demonstrated ballistic graphene Josephson junction in the overdamped regime. 



Graphene is made using the exfoliation method\citep{Novoselov10451} and is encapsulated in hexagonal boron-nitride (hBN) using the “pick-up” method \citep{Wang614}. Using CHF$_{3}$/O$_{2}$ plasma, the hBN-Grahpene-hBN stack is etched thorough in order to make quasi-one dimensional electrical contacts with superconducting electrodes\citep{Our_Science}.  Molybdenum-Rhenium(MoRe) alloy electrodes are deposited onto the device using DC sputtering with the approximate thickness of $100-120$nm. The bonding pads and thin metal leads making contact to MoRe are made of Cr/Au ($5$nm/$110$nm). Here, we present data on the device of length $L=500\mathrm{nm}$ (the distance between MoRe contacts), and the width $W=3\mu\mathrm{m}$ (see supplementary).

\begin{figure}[ht]
\includegraphics[width=0.5\textwidth]{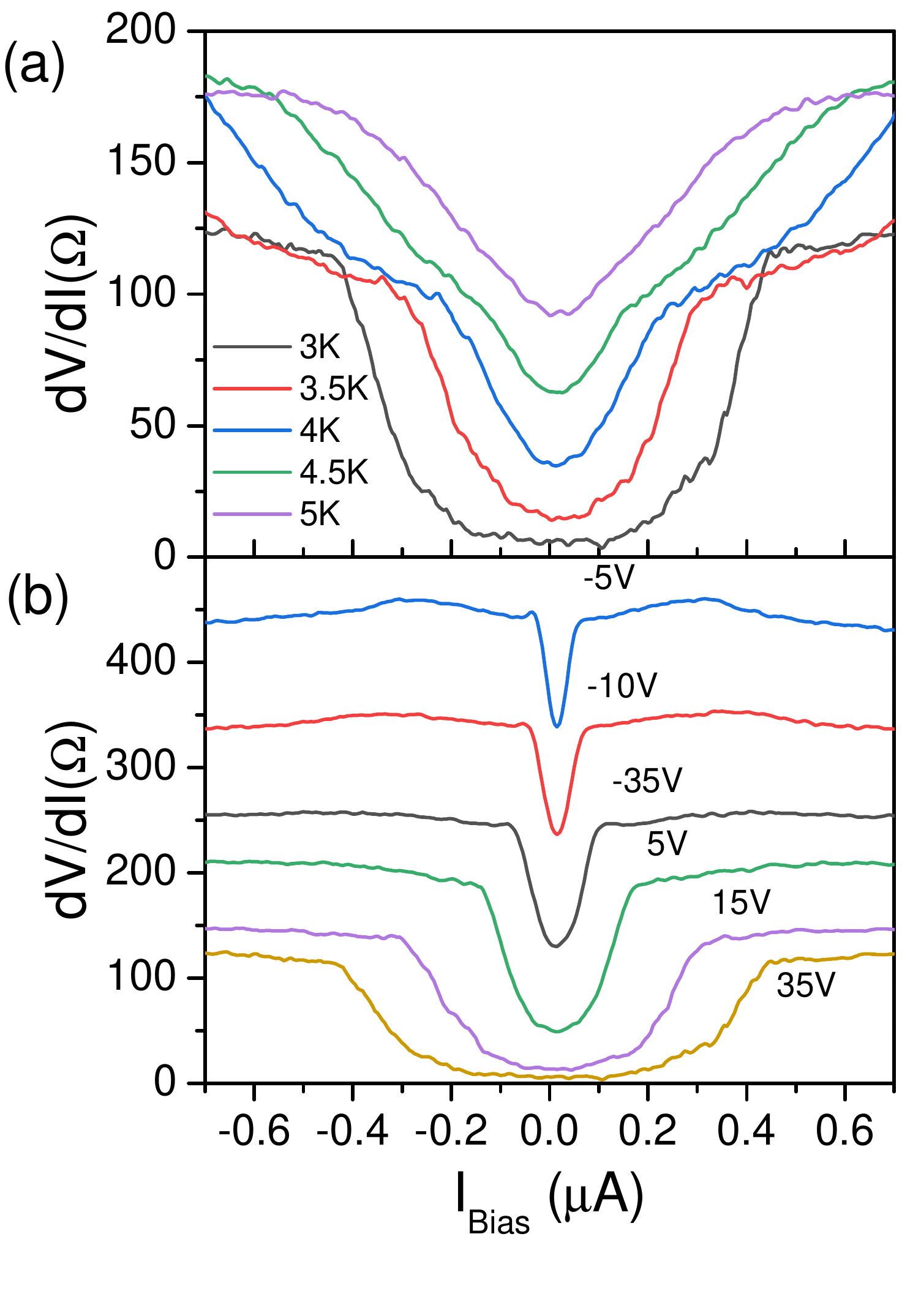}

\caption{\label{fig:epsart} Differential resistance $dV/dI$ as function of DC bias current $I_{Bias}$. The current is swept from negative to positive. Panel (a) shows $dV/dI$ curves for different temperatures. The gate voltage here is set to 35V. Panel (b) shows $dV/dI$ curves for different applied gate voltages. The temperature here is 3K. For both of the figures we can notice that there is no observable hysteresis between the switching and retrapping current. All curves feature a measurable resistance even at zero bias current. This zero-bias resistance $R_0$ arises from the phase diffusion mechanism.}
\end{figure}

\begin{figure}[h!]
\includegraphics[width=0.5\textwidth]{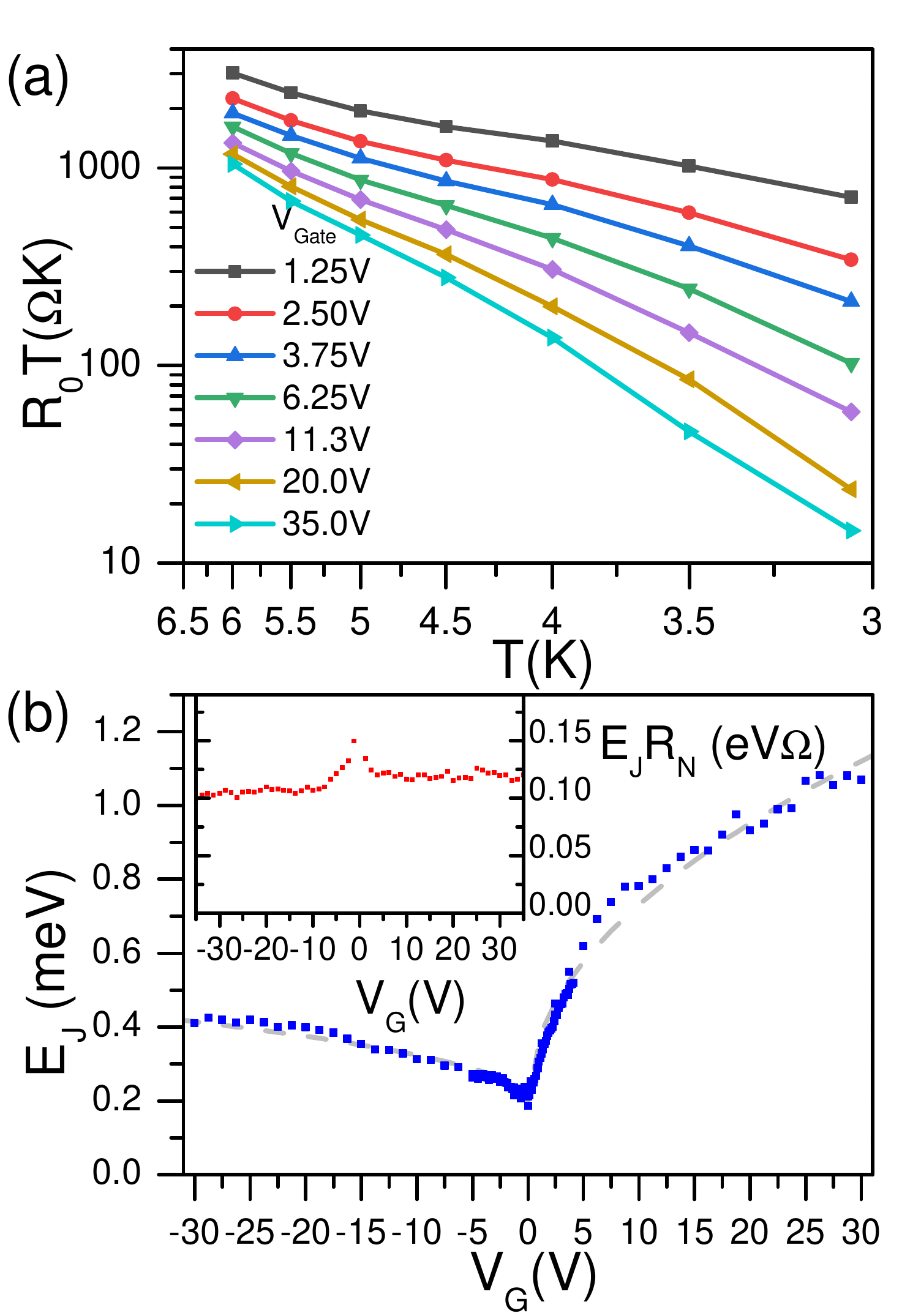}

\caption{\label{fig:epsart} (a) The product of temperature and the zero-bias resistance $R_0\cdot T$ as a function of inverse temperature $1/T$ shown for several different gate voltages plotted on a semi log scale. The linear dependence of the curves shown here confirms that the resistance $R_0$ arises from the phase diffusion mechanism. (b)The Josephson energy $E_J$ versus gate voltage $V_G$. Here $E_J$ is calculated from the slope of the curves shown in panel (a). The dashed lines represent a square root relationship bewteen $E_J$ and $V_G$ (with an offset), the expected relationship for single layer graphene Josephson junctions in the long ballistic regime. (Inset) The product of the Josephson energy and normal resistance $E_J \cdot R_N$. For ballistic, single layer graphene Josephson junctions, this product is expected to be constant with respect to gate voltage $V_G$. The discrepancy between the electron and hole conduction regime comes from the negative contribution of the contact interface transparency which has been found to suppress the critical current $I_C$. }
\end{figure}

The device is measured in a home-made cryocooler with a base temperature of ~2.5K, which is isolated via both a heat shield and RC filters placed at the low temperature stage. Josephson junction resistance is measured using the lockin method with a four-probe geometry. The junction is biased by a variable DC current with a small AC excitation of $5$nA. The voltage across the junction is amplified using a custom differential pre-amp prior to being fed into the lockin.  The gate voltage applied to the back of the 300nm SiO$_{2}$ oxide layer is used to control the carrier density of graphene. Figure 1 presents the differential resistance $dV/dI$ as a function of applied DC bias current $I_{Bias}$. All the curves show two transition points, as the bias current is swept from a large negative value to a large positive value. The absolute value of the current on the negative side below which the junction becomes superconducting is the retrapping current $I_R$, i.e. $|I_{Bias}|=I_R$. On the positive side, the junction transitions from the superconducting to the normal state at the switching current $I_S$\citep{tinkham}. Figure 1(a) shows resistance versus $I_{Bias}$ for different temperatures with the backgate voltage set to 35V. Figure 1(b) shows the resistance versus bias current for different gate voltages taken at 3K. In both cases, the switching and retrapping currents $I_S$ and $I_R$ follow the expected trends: falling exponentially with increasing temperature, and increasing with gate voltage away from the Dirac point with the hole conduction regime exhibiting a suppressed critical current due to the effects of contact doping\citep{Phase_Diff,Ke2016,Borzenets_Long_Ballistic}. Moreover, the trend of $I_S$ with respect to temperature $T$ follows that expected for ballistic Josephson junctions in the long junction regime (See Supplementary)\citep{Borzenets_Long_Ballistic}.

The vast majority of previously reported graphene Josephson junctions, exhibit hysteresis between the switching and retrapping currents, with $I_S>I_R$ even for temperatures above 3K. While certain works attribute this hysteresis to self-heating of the junction\citep{Borzenets_Phonon, PhysRevLett.101.067002}, it has also been shown that most of the graphene Josephson junctions exhibit underdamped behavior\citep{Phase_Diff, Ke2016, BenShalom2016, Borzenets_Long_Ballistic}. (Graphene-based Josephson junctions a typically overdamped due to the large capacitance caused by the presence of the backgate.)  Here, however, for all tested gates and temperatures, we do not observe a difference between $I_S$ and $I_R$. This leads to an initial indication that the device may be overdamped. 


We now further investigate the junction dynamics via the characterization of device behavior in the phase diffusion regime. Phase diffusion manifests itself as an observable non-zero resistance below the critical current (even at $I_{Bias}=0$), arising from phase slips that are caused by thermal noise. The rate of these phase slips down the prototypical titled washboard potential, and therefore, the measured zero-bias resistance is governed by the junction dynamics which dictate the energy dissipation rate\citep{tinkham}. Indeed, we are able to observe a measurable resistance in our devices, even at zero bias, and down to 3K in temperature. We define the measured zero-bias differential resistance as R$_0$. In order to confirm that $R_0$ arises from the phase diffusion mechanism, we study the evolution of this resistance with respect to temperature. Regardless of the junction damping dynamics the trend behavior of $R_0$ with respect to temperature should have the following dependence: \citep{tinkham, Phase_Diff,PhysRevB.50.395,PhysRevLett.63.1507,PhysRevB.42.9903,JETP.8.113,JETP.28.1272, Ambegaokar}:
\begin{equation}
R_{0} (T) \propto \frac{2E_ {J}}{k_ {B} T}e^{-2E_ {J} / k_ {B} T} 
\end{equation}

Here, $E_J= \hbar I_C/2e$ is the Josephson energy. (The above exponential dependence holds when $2E_ {J} / k_ {B} T >1$ \citep{tinkham}.) Reworking equation (1), we can arrive at the proportionality relationship:  $\log{(R_{0}(T) \cdot T)} \propto -2E_{J}/k_{B}T$. In Figure 2a we plot the value $R_0 \cdot T$ versus inverse temperature  $T$ on a semi-log scale. Indeed, we find that the relationship is nearly linear, consistent with theory. From here, knowing the temperature, we can extract $E_J$ from the slope of the curves.  The fitted Josephson energy versus gate voltage is plotted in Figure 2b. For single layer graphene Josephson junctions governed by ballistic transport it is expected that value $E_{J}\cdot R_{N}$ is constant with respect to $V_G$\citep{Borzenets_Long_Ballistic}; and we find not only that indeed this is the case (see Figure 2 inset), but that this value matched well with the expected energy scale extracted from the trend of $I_S$ vs $T$ (see supplementary).  Here $R_N$ is the normal resistance, i.e. the resistance of the junction when it is in normal state.  

Having found the Josephson energy, we can determine the last parameter governing $R_0$. This final parameter is different depending on the damping dynamics of the junction. Previous theoretical works have defined three different regimes: For overdamped junctions, the governing parameter is $R_N$, the normal resistance\citep{JETP.28.1272, Ambegaokar}. For underdamped junctions,  $R_0$ depends on the plasma frequency $\omega _p \propto \sqrt{E_J / C}$\citep{PhysRevB.42.9903}. (Here $C$ is the capacitance of the junction.) Finally, if the junction is underdamped at low frequencies, but becomes overdamped at the plasma frequency (due to being shorted by the environment), we have $R_0\propto Z_0$\citep{PhysRevB.50.395}. Here $Z_0$ is the real part of the high frequency impedance caused by the junction's environment \citep{PhysRevB.50.395}. We find that analyzing our devices as if they were underdamped, or damped by the environment does not produce a good match between measured data and theoretical expectation. (See supplementary).

\begin{figure}[ht!]
\includegraphics[width=0.45\textwidth]{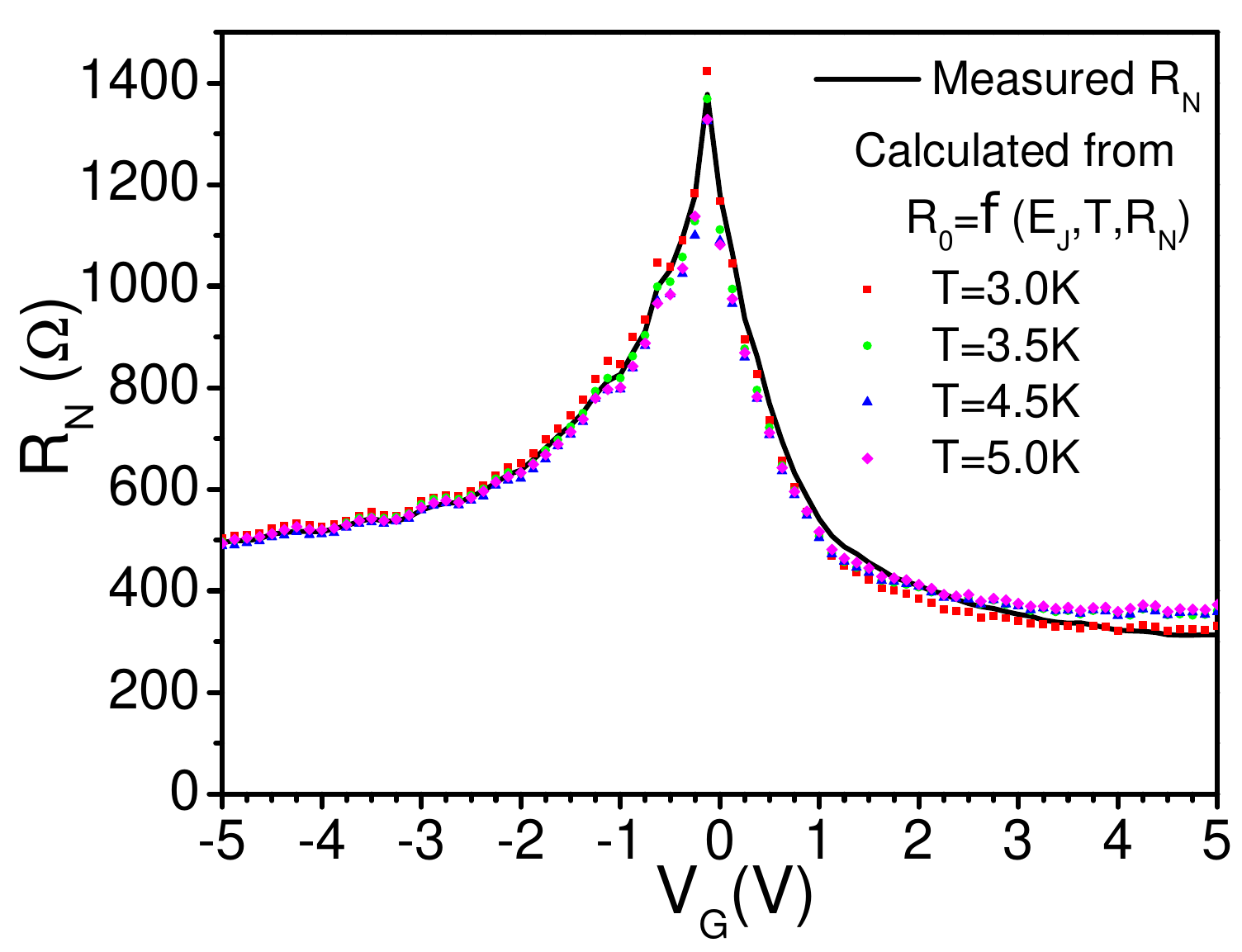}

\caption{\label{fig:epsart}  Gate voltage dependence of measured normal resistance $R_N$ (black line) compared to the resistance calculated from the theoretically predicted relationship between $R_N$ and $R_0$: $R_N = R_0/[I_0(E_J/k_BT)]^{-2}$ . (I$_0$ is the modified Bessel function.) The measured $R_N$ here is taken at 3.5K. The data and calculated result match well, supporting the claim that the measured device is and overdamped junction.}
\end{figure}

Now, we confirm that our devices are indeed in the overdamped regime by comparing the measured normal resistance $R_N$ with that back calculated from the the zero-bias resistance $R_0$. Following the full expression in Ref. \citep{Ambegaokar} we have:
\begin{equation}
\lim_{I_{Bias} \to 0} \frac{V/I_{Bias}}{R_N}=[I_{0}(\frac{1}{2} \gamma)]^{-2}
\end{equation}
Here, I$_0$ is the modified Bessel function, and $\gamma =I_c \hbar/ek_B T$. ($V$ is the voltage measured accross the junction.) For $I_{Bias}$ approaching zero, the equation simplifies to $R_N=R_0/[I_0(E_J/k_BT)]^{-2}$. Figure 3 shows the measured normal resistance $R_N$ versus gate voltage $V_G$ plotted together with the resistance calculated from Equation 2 for different temperatures. It can be seen that we have a good match between the measured and the theoretical result, in particular for high values of $R_N$.  

The damping of the junction is typically determined by the quality factor $Q$, with $Q=R_N(2eI_CC/\hbar)^{1/2}=\frac{2e}{\hbar}R_N\sqrt{E_J C}$. ($C$ being the capacitance of the junction.) A  $Q < 0.85$ results in an overdamped junction\citep{Likharev}. However, the difficulty in calculating an accurate quality factor $Q$ is two fold: One, the measured critical current $I_C$ has been consistently less than the value predicted by theory, leading to the question of whether the true critical current is being measured. Two, it is unclear which areas of the device play an active role in the junction capacitance $C$. (Previous works have suggested that the capacitance all the way up to the capacitive coupling between the device bonding pads plays a significant role. Thus the junction capacitance is always several orders of magnitude larger compared to that expected from the dimensions of the SNS region itself.) Assuming that the junction capacitance includes the contribution of the $100\times 100\mu\mathrm{m}$ bonding pads that couple to each other via the backgate below the $300\mathrm{nm}$ thick $\mathrm{SiO_2}$ layer, we arrive at $C\approx 600\mathrm{fF}$. Taking $E_J=0.3\mathrm{meV}$ and $R_N=100\Omega$, we calculate a quality factor $Q=1.6$: the underdamped regime. Hence, we are led to conclude that the tested junction has been sufficiently isolated from the capacitive contribution of the bonding pads. Indeed, we estimate that the interface between MoRe superconducting contacts and the gold leads suffers from a contact resistance in the order of tens of $\Omega$, which is sufficient to reduce the quality factor below $1$\cite{Vion, Borzenets_Phonon}.
  
In conclusion, we have investigated the phase diffusion regime in hBN encapsulated graphene Josephson junction governed by ballistic electron transport. The observed trend of the measured zero-bias resistance $R_0$ with respect to carrier concentration conforms well to theory describing phase diffusion in an overdamped junction regime. This is the first conclusive confirmation of overdamped behavior in graphene-based Josephson junction. We attribute this behavior to effective isolation of the Josephson junction from the capacitive contribution of the bonding pads. The isolation arises from a resistive connection with the device layout.

\begin{acknowledgements}
J.T., A.Sh., and I.V.B. acknowledge CityU New Research Initiatives/Infrastructure Support from Central (APRC): 9610395, and the Hong Kong Research Grants Council Projects: (ECS) 2301818, (GRF) 11303619. Lithographic fabrication and characterization of the samples performed by  E.G.A., and A.S. were supported by the Division of Materials Sciences and Engineering, Office of Basic Energy Sciences, U.S. Department of Energy, under Award DE-SC0002765. 

\end{acknowledgements}


\bibliography{Overdamped_Phase_Diffusion_4_2_arXiv}

\clearpage
\newpage

\onecolumngrid
\begin{center}
\textbf{\large Supplementary to: Overdamped Phase Diffusion in hBN Encapsulated Graphene Josephson Junctions}
\end{center}

\beginsupplement

\section*{ Device Design and Characterization }
The optical image of the device presented in the main text is shown in Figure S1a. Graphene is made using the exfoliation method\citep{Novoselov10451}, prior to encapsulation in hexagonal boron-nitride (hBN) the graphene is verified to be single layer using Raman spectroscopy\cite{Raman}. The measured Raman spectrum is shown in Figure S1b. An AFM image is taken of the completed hBN-graphene-hBN stack (not shown), and an area free of bubbles and defects is chosen for further processing. Using CHF$_{3}$/O$_{2}$ plasma, the hBN-Grahpene-hBN stack is etched thorough in order to make quasi-one dimensional electrical contacts with superconducting electrodes\citep{Our_Science}.  Molybdenum-Rhenium(MoRe) alloy electrodes are deposited onto the device using DC sputtering with the approximate thickness of $100-120$nm. The MoRe contacts define the Josephson junction dimensions. However, the MoRe leads are terminated $\sim50\mathrm{\mu m}$ past the active area of the device. (Green arrow in Figure S1a). The bonding pads and thin metal leads making contact to MoRe are made of Cr/Au ($5$nm/$110$nm). This interface between the MoRe and the Cr/Au portion of the contacts results in a resistance that isolates the Josephson junction from the effect of the parasitic capacitance between the bonding pads\citep{Vion, Borzenets_Phonon}

\begin{figure}[ht!]
\includegraphics[width=0.5\textwidth]{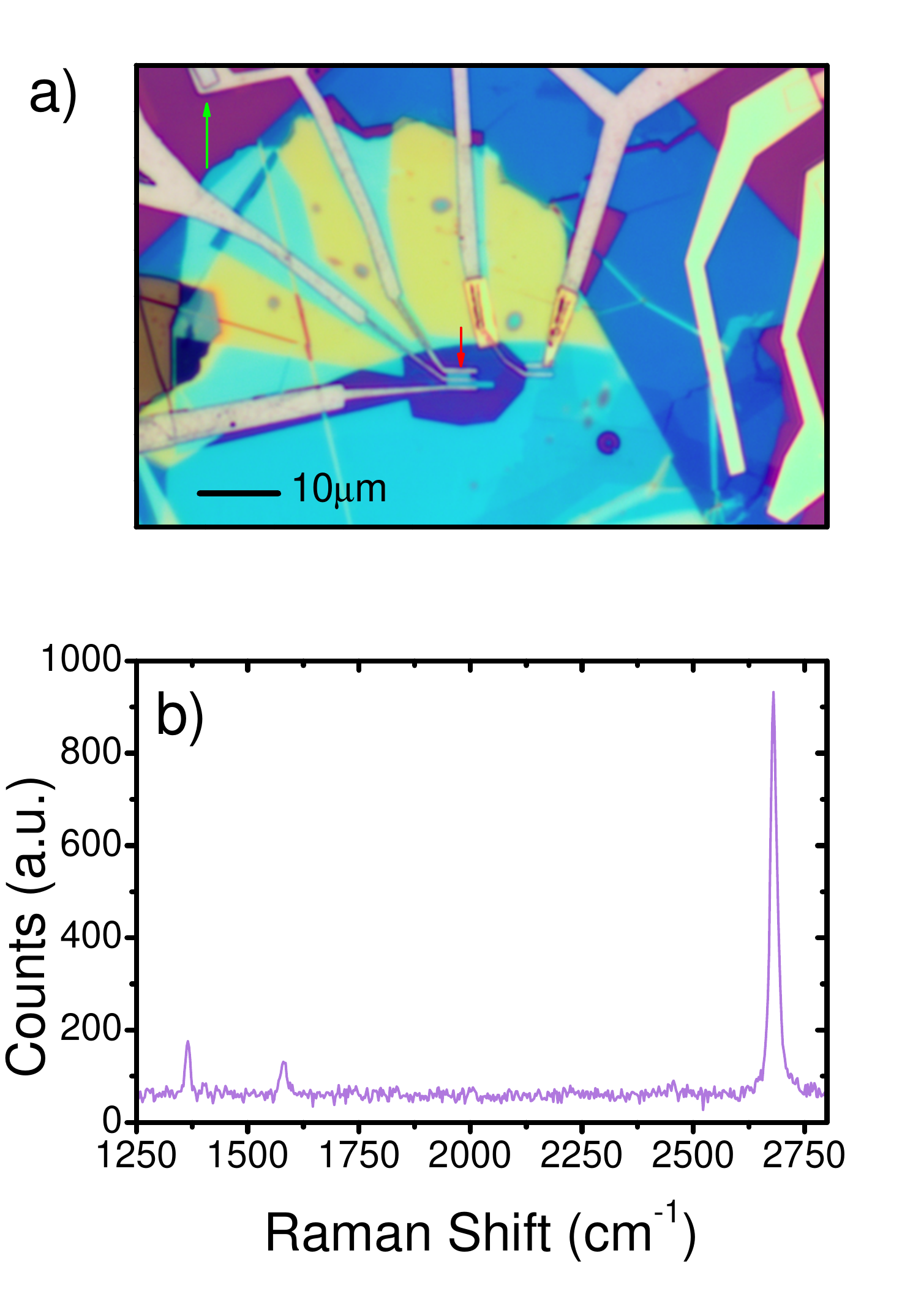}
\caption{\label{fig:epsart}  a)Optical image of the device presented in the main text. (Scale bar presented for reference). Graphene encapsulated in hexagonal boron-nitride (hBN) acts as the normal metal portion of the device. The superconductor is made from MoRe alloy, with the superconducting leads defining a junction of $500$nm length. The MoRe leads terminate  $\sim50\mathrm{\mu m}$ past the active area of the device. The bonding pads and leads connecting to the MoRe region are made of Cr/Au, thickness  $5$nm/$110$nm).  An example of the transition between MoRe and Cr/Au is highlighted by the green arrow. The red arrow indicated the junction presented in the main text. b)The Raman spectrum of the graphene region used in the device. }
\end{figure}

\begin{figure}[ht!]
\includegraphics[width=0.5\textwidth]{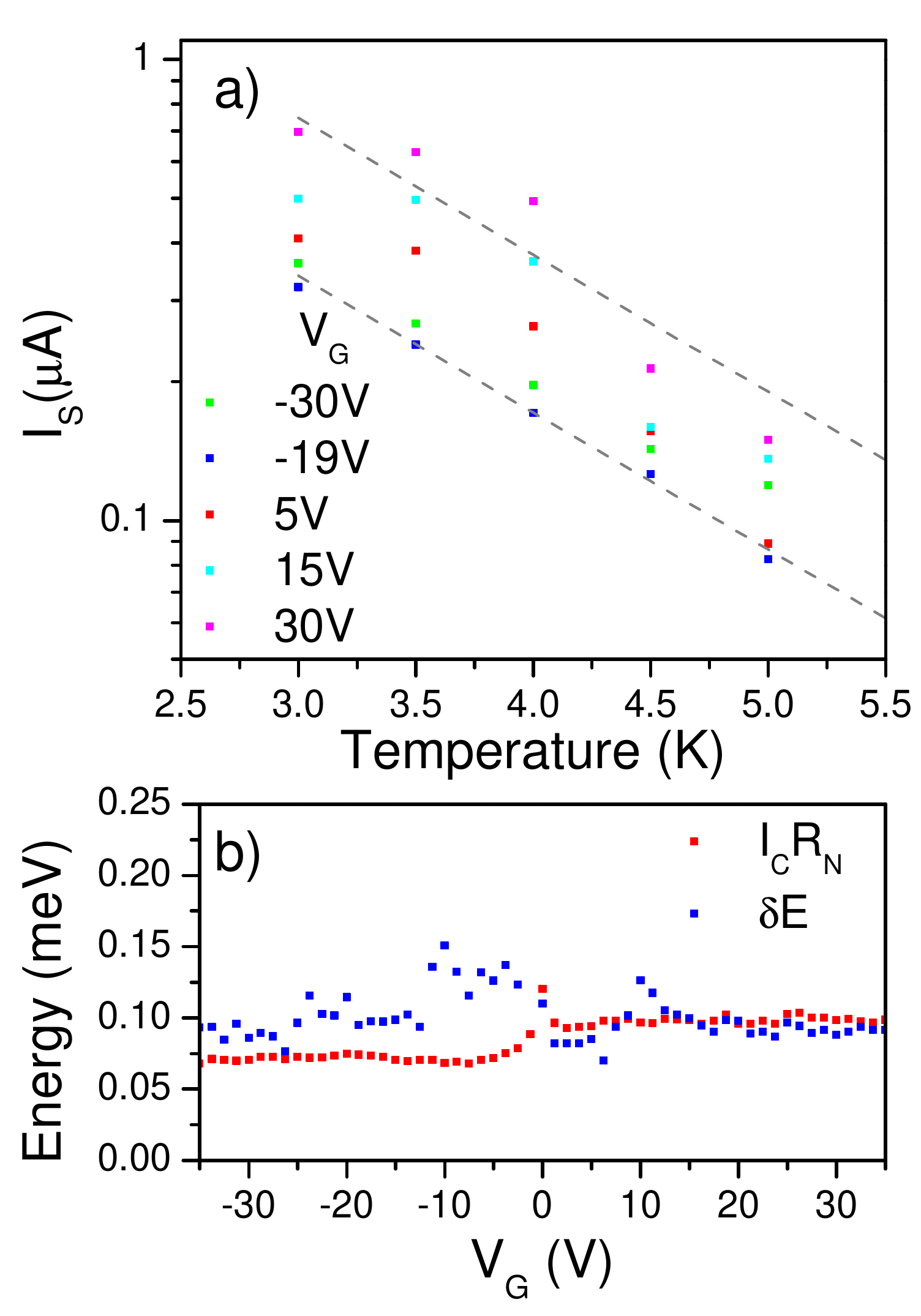}

\caption{\label{fig:epsart}  a)Switching current $I_S$ versus temperature $T$ plotted on a semi-logarithmic scale. Plots for several values of gate voltage are shown. The plot shows a nearly linear decay with respect to $T$, from the slope of the decay one can fit the relevant energy scale $\delta E$. The fit results are plotted in panel b), and also as dashed guide-lines in panel a). b)The fitted energy scale $\delta E$ with respect to gate voltage $V_G$. A $\delta E$ that is independent of gate is expected for single layer graphene. Alongside, is plotted the value $I_C R_N$ (critical current times normal resistance). The two sets of data are very similar, as is expected.  }
\end{figure}

\section*{ Switching current $I_S$ dependence on Temperature $T$}
In Josephson junctions, the measured switching current $I_S$ is suppressed from the expected maximum critical current $I_C$, with $I_S$ found to be decreasing with increasing temperature $T$ \citep{tinkham}. Measuring the trend of $I_S$ with respect to temperature $T$, one can determine whether the junction is in the diffusive\citep{Ke2016,Dubos2001,Dubos2001_2} or ballistic transport regime\citep{Kulik,Bardeen, Svidzinski1, Svidzinski2,Beenakker92, Beenakker91, Lee,Borzenets_Long_Ballistic}. Additionally, by sweeping gate voltage $V_G$,  one can also determine whether the junction is made of single layer or multi-layer graphene. For the case of a Josephson junction governed by ballistic electron transport, and with junction length $L\gtrapprox \xi$ (the superconducting coherence length) we expect that $I_S(T)\propto \exp(- k_B T/\delta E)$\citep{Kulik,Bardeen, Svidzinski1,Svidzinski2,Borzenets_Long_Ballistic}.  The value $\delta E\approx\hbar v_F/2\pi L$ is related to the Andreev Bound States energy level spacing ($E_0 = \pi\hbar v_F/L$)\cite{Kulik,Bardeen,book,tinkham,Golubov}. Here $v_F$ is the the Fermi velocity; and in single layer graphene $v_F$ is a constant with respect to carrier density. Therefore, $\delta E$ is expected to be independent of the applied gate voltage $V_G$ (as long as the junction remains ballistic).
The trend of the switching current $I_S$ versus temperature $T$ can be seen in Figure S2a, plotted on a semi-log scale for several values of $V_G$. A nearly linear decay can be observed. Fitting the data to the above exponential dependence, we can extract the value of $\delta E$.The guidelines of the fit result are shown as dashed lines in Figure S2a, while the fitted value of $\delta E$ is presented in Figure S2b. One can see that $\delta E$ is independent with respect to gate voltage $V_G$ (aside from the expected deviations close to the Dirac point).  Moreover, the extracted value is consistent with the designed length $L$ of the junction\citep{Borzenets_Long_Ballistic}. 
As the temperature approaches zero, the switching current approaches the critical current with $I_C R_N = C \delta E$. (Here $R_N$ is the normal resistance.)
It has been found empirically, that the dimensionless proportionality constant $C$ is $\approx 1-2$, and is suppressed by the coefficient of transmission $\tau$ across the junction\citep{Borzenets_Long_Ballistic}.  In Figure S2b, alongside $\delta E$, we plot the value of $I_C R_N$. The critical current $I_C$ is back calculated from the fitted Josephson energy $E_J$ in the main text.   Indeed, we find a good match between the two values. This suggests that our device is single layer graphene, and is governed by ballistic electron transport. Moreover, that our fitted Josephson energy is close to the actual $E_J$ of the device.

\section*{ Comparisson to Underdamped and Damped by Environment regimes }
The main text presents the analysis of phase diffusion in the overdamped regime. Here we follow the analyses as if the Josephson junction was underdamped, or damped by the environment at the plasma frequency. In following previous works,  define the prefactor to the exponential in the main text Equation 1 as a variable  $R_0^{\prime}$ with $R_0^{\prime}\equiv R_0 e^{(2E_J/k_B T) }$ \citep{Phase_Diff}.  
For underdamped junctions,  $R_0^{\prime} \sim (h/e^2) \hbar \omega _p/k_B T$ \citep{PhysRevB.42.9903}, where $\omega _p \propto \sqrt{E_J / C}$ is the plasma frequency. Note that $C$ is the capacitance of the junction, and is a constant.) However, if the junction is underdamped at low frequencies, but becomes overdamped at the plasma frequency (due to being shorted by the environment), we have $R_0^{\prime}=2\pi Z_0 E_J /k_B T$\citep{PhysRevB.50.395}. Here $Z_0$ is the real part of the high frequency impedance caused by the junction's environment \citep{PhysRevB.50.395}. Typically $Z_0$ is found to be $\sim200-250\mathrm{\Omega}$, and can also be treated as a constant. (Note, that the above relationship holds only for $E_J>k_bT$\cite{PhysRevB.50.395}).
The measured  $R_0^{\prime}$ versus the Josephson energy $E_J$ is shown in Figure S3 (data taken at 4.0K). Alongside, we plot the best fitting results obtained for each of the two cases mentioned above. Note that the trend of the measured result is super-linear; while the expected trend for the underdamped case is a square root dependence, and a linear trend is expected in the damped by the environment case. Clearly, the above two analyses do not apply. It is important to note that Ref.  \onlinecite{Phase_Diff}, erroneously used the same analysis via the variable $R_0^{\prime}$ for the overdamped case. This is not correct as the exponential dependence only approximates the slope of the trend with respect to temperature, not the absolute value\citep{tinkham, Ambegaokar}. Instead, the full expression from Ref. \onlinecite{Ambegaokar} is to be used; as was done in the main text. 

\begin{figure}[ht!]
\includegraphics[width=0.5\textwidth]{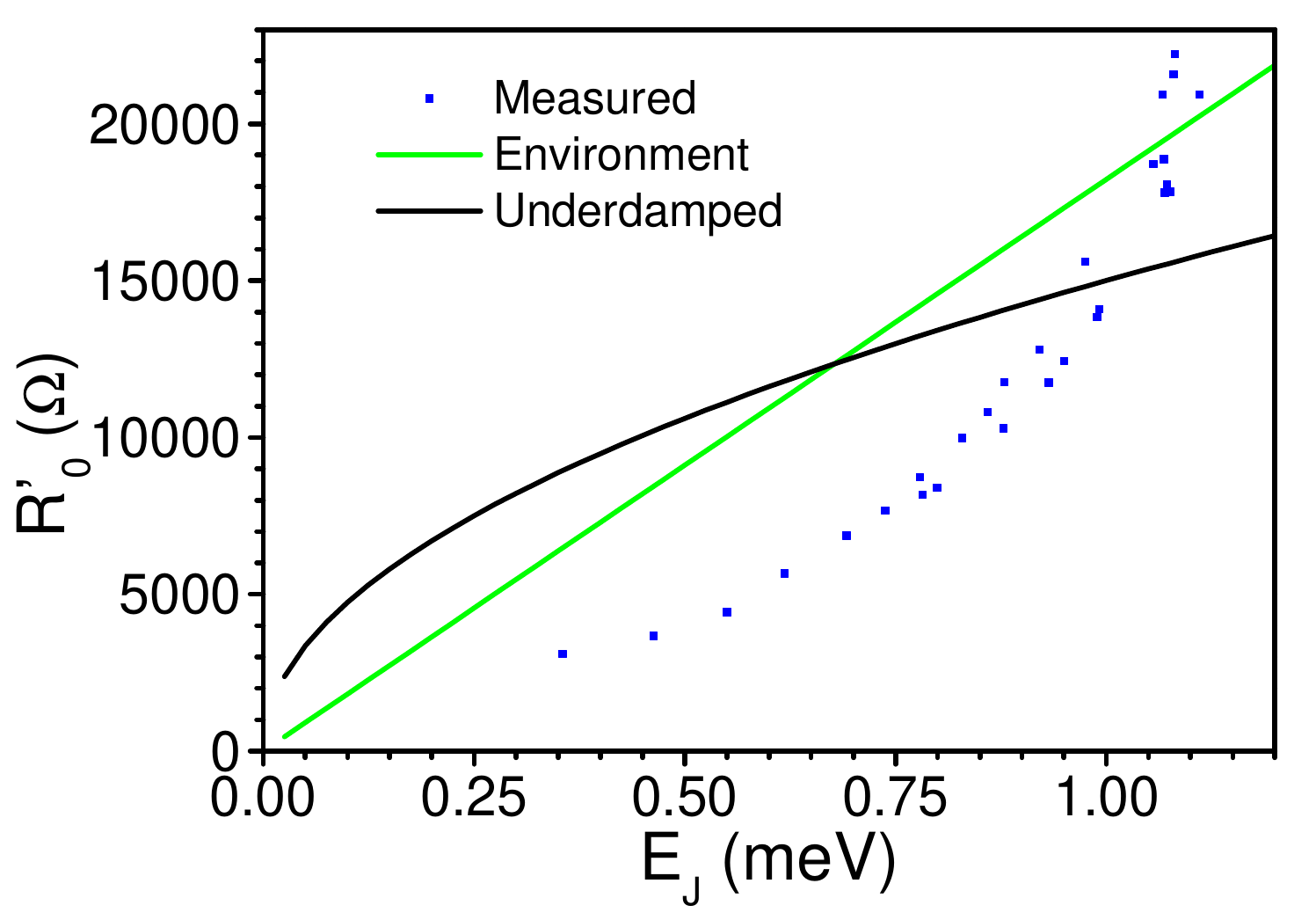}

\caption{\label{fig:epsart}  The variable $R_0^{\prime}$ versus the Josephson energy $E_J$. The blue scatter data represents results obtained from the measured data. The black and green lines represent best fit lines for the theoretical expectation in the case the Josephson junction is underdamped or damped by the environment (respectively). The mismatch between the measured data and theoretical predictions suggest than neither of these cases apply.  }
\end{figure}

\end{document}